\documentclass{elsart}
\input{psfig}
\usepackage{natbib}
\begin{document}

\

\vskip4cm

\Large
\centerline{\bf ~~~  THE HUBBLE CONSTANT}
\centerline{\bf   AND THE EXPANSION AGE OF THE UNIVERSE}
\bigskip
\bigskip

\centerline{~~~Wendy L. Freedman$^1$}
\normalsize 
\bigskip
\vskip3cm
\vskip3cm

\noindent
$^1$ Carnegie Observatories, 813 Santa Barbara St., Pasadena, CA 91101.

\bigskip
\bigskip

\noindent
To be published in the {\it David Schramm Memorial Volume},  Physics
Reports,  Elsevier, 2000, in press.

\vfill\eject

\medskip

\runauthor{Freedman}
\begin{frontmatter}
\title{The Hubble Constant and the Expansion Age of the Universe}
\author[wlf]{Wendy L. Freedman}

\address[wlf]{Carnegie Observatories, Pasadena CA}


\begin{abstract}

The Hubble constant, which measures  the expansion rate, together with
the total  energy density  of  the Universe,   sets  the size of   the
observable Universe, its age, and its  radius of curvature.  Excellent
progress has been  made recently toward  the measurement of the Hubble
constant: a number  of different methods  for measuring distances have
been developed and refined, and a primary project  of the Hubble Space
Telescope has     been   the     accurate    calibration of       this
difficult-to-measure  parameter.   The   recent   progress  in   these
measurements is summarized, and areas where further work is needed are
discussed. Currently,  for  a  wide  range  of  possible  cosmological
models, the Universe appears to  have a kinematic  age less than about
14$\pm$2 billion years.  Combined  with  current estimates of  stellar
ages, the results  favor  a low--matter--density universe.    They are
consistent with  either an  open universe, or  a flat  universe with a
non-zero value of the cosmological constant.

\end{abstract}

\end{frontmatter}

\typeout{SET RUN AUTHOR to \@runauthor}


\section{Introduction}

The Hubble constant (H$_0$) is one of the most important parameters in
Big Bang  cosmology:  the square of the   Hubble constant relates  the
total energy  density of the  Universe to  its geometry \cite{kt,pea}.
H$_0$  enters  in a practical  way  into  many  cosmological and other
astrophysical calculations:  together with  the energy density  of the
Universe,   it  sets the   age  of the Universe,   t,  the size of the
observable Universe (R$_{obs}$ =  ct), and its  radius of curvature ($
R_{curv}  = {c \over  H_0} { {(\Omega  - 1)} \over  k }^{-{1 \over 2}}
$).  The  density of light   elements (H,  D,  $^3$He,  $^4$He and Li)
synthesized after   the Big Bang also depends   on the expansion rate.
These limits on the density of baryonic matter can then be used to
set limits on the amount of non-baryonic matter in  the Universe.  The
determination of numerous physical properties  of galaxies and quasars
(mass, luminosity, energy density) all require knowledge of the Hubble
constant.


Primarily  as    a result  of    new  instrumentation at  ground-based
telescopes, and most recently with the successful refurbishment of the
Hubble Space Telescope  (HST), the extragalactic distance scale  field
has been evolving  at  a rapid pace.    Still, until very recently,  a
factor-of-two uncertainty in  the value of  H$_0$ has persisted for  a
variety of  reasons \cite{jacoby92,  wlfprinceton}.  Since the 1980's,
linear detectors, replacing   photographic plates,  have enabled  much
higher accuracy measurements, corrections for the effects of dust, and
measurements to much greater  distances, all combining to increase the
precision  in   the relative  distances   to  galaxies. Prior  to HST,
however, very few galaxies were close enough to allow the discovery of
Cepheid  variables, upon  which    the absolute  calibration  of   the
extragalactic  distance     scale   largely  rests     \cite    {mf91,
fmtrimble}. 

In the  following sections  I  summarize  how the Hubble   constant is
measured in practice,  and the  problems  encountered in  doing so.  I
describe in general how to  measure distances, list both the strengths
and  weaknesses of various  methods for  measuring distances, and then
discuss  the factors that  affect the determination of true, expansion
velocities.  In addition, I briefly review the method, the results and
the uncertainties for the determination Cepheid distances to galaxies,
and recent results from the Hubble  Space Telescope (HST) by the H$_0$
Key Project and  other groups. I then give  the implications of  these
results for cosmology, and  compare these ``local'' results to methods
that can be applied directly  at high redshifts.  Finally, I highlight
areas where future work would be profitable.

\section{Measuring the Hubble Constant}

Determination of the Hubble constant is extremely simple in principle:
measure the recession   velocities and the  distances   to galaxies at
sufficiently  large distances where  deviations from the smooth Hubble
expansion are small, and the Hubble  constant follows immediately from
the slope of the correlation between  velocity and distance.  However,
progress in  measuring H$_0$ has been limited  by the fact  that there
exist few  methods  for measuring  distances that  satisfy  many basic
criteria.    Ideally,  a  distance   indicator should   be based  upon
well-understood physics, operate well  out into the smooth Hubble flow
(velocity-distances greater than $\sim$10,000 km/sec), be applied to a
statistically   significant     sample  of  objects,   be  empirically
established to have high  internal accuracy, and most importantly,  be
demonstrated  empirically to be free  of systematic errors.  The above
list of criteria applies equally well to classical distance indicators
as to  other physical methods  (in the  latter case, for  example, the
Sunyaev Zel'dovich effect or gravitational lenses).

Historically,  measuring  accurate   extragalactic distances has  been
enormously  difficult;  in  retrospect, the  difficulties   have  been
underestimated and systematic errors have dominated.  And still today,
the  critical remaining issue is to  identify and reduce any remaining
sources of systematic error.   At the present  time, an ideal distance
indicator or other  method meeting all of the  above criteria does not
exist, and measurement of H$_0$ as high as 1$\%$ accuracy is clearly a
goal  for the  future.   However, as described  below,  an accuracy of
H$_0$ to 10\% has now likely been reached.

\section{\bf Recession Velocities} 

Since the velocity  of recession of a  galaxy  is proportional  to its
distance (Hubble's law), the farther that distance measurements can be
made, the  smaller the proportional impact  of peculiar motions on the
expansion velocities. For a galaxy  or cluster at a recession velocity
of 10,000  km/sec,  the impact  of a  peculiar  motion of  300  km/sec
\cite{giovanelli98} is 3\% on H$_0$ for that object.  This uncertainty
is reduced by observing a number of objects, well-distributed over the
sky, so that  such motions can be  averaged out.   Moreover, given the
overall mass distribution locally, a  correction for peculiar  motions
can be  applied to the velocities (over  and above corrections for the
Earth's, Sun and our Milky Way's motion in the Local Group).  For type
Ia  supernovae,  the distant   indicator which  currently  extends the
farthest (v $\sim$ 30,000 km/sec), the effects of peculiar motions are
a small fraction of the overall error budget.

\section{Distances to Galaxies}

In astronomy most length  scales  cannot be  measured directly --  the
size scales,  especially  in  a cosmological   context are too   vast.
Direct trigonometric parallaxes (using the Earth's orbit as a baseline
for triangulation) can be measured for the  nearest stars in our Milky
Way galaxy, but this technique currently can  be applied reliably only
for relatively nearby stars within our own Galaxy.  
More distant  stars in our  Galaxy and then  extragalactic objects
require other, more indirect indicators of distance.

In general, the   most  common  means for  estimating    extragalactic
distances  make use of the inverse  square radiation  law.  If objects
can  be identified whose luminosities  are either constant (``standard
candles''),  or perhaps related to a   quantity that is independent of
distance (for example, period  of oscillation, rotation rate, velocity
dispersion,   or color)  then   given  an absolute calibration,  their
distances   can be   gauged.     The  ``standard candles''   must   be
independently   calibrated to  absolute physical  units  so that  true
distances (in units  of   megaparsecs, where  1 Mpc =   3.09  $\times$
10$^{22}$ m) can  be determined.  Ultimately, these  calibrations tie
back to   geometric parallax distances.   Alternatively  a  ``standard
ruler'' can  be used, making use of  the fact that physical dimensions
scale inversely  as the   distance.    Several methods for   measuring
distances to galaxies are summarized below.

\subsection{Cepheid variables}

Primary amongst the  distance  indicators are the  Cepheid  variables,
stars  whose outer atmospheres  pulsate regularly with periods ranging
from  2 to about 100 days.  Cepheids are bright, young stars, abundant
in  nearby spiral and irregular galaxies.   The  underlying physics of
the   pulsation mechanism is  simple and  has been studied extensively
\cite{cox}.  Empirically it has   been established that the period  of
pulsation (a quantity independent of distance) is very well correlated
with  the intrinsic  luminosity of  the  star.  The dispersion in  the
Cepheid period-luminosity   relation in     the I   band   ($\sim$8000
Angstroms)  amounts to about   20\% in  luminosity.  From  the inverse
square law,  this corresponds to an  uncertainty of about  10\% in the
distance for a single  Cepheid.   With a sample  of  25 Cepheids in  a
galaxy,  a  statistical uncertainty of  about  2\% in  distance can be
achieved.  Hence, Cepheids  provide an  excellent means of  estimating
distances to resolved spiral galaxies.  I return in  \S\S 4.7, 5.2 and
6.1 to   a discussion of the largest   remaining uncertainties  in the
Cepheid distance scale.

The reach   of Cepheid variables  as  distance  indicators is limited.
With available  instrumentation, for  distances beyond  20 Mpc or  so,
brighter   objects than  ordinary stars  are   required; for  example,
measurements   of luminous supernovae   or  the luminosities of entire
galaxies.  Implementation of  these secondary methods are  now briefly
described in turn.


\subsection{Type Ia Supernovae}

Perhaps the most promising of the cosmological distance indicators are
the luminous   supernovae classified as type   Ia.  Type Ia supernovae
show no hydrogen in their spectra, and are believed to result from the
explosion  of a carbon-oxygen white  dwarf which burns into $^{56}$Ni.
\cite{livio}.   These objects have  luminosities comparable  to entire
galaxies   of moderate   luminosity, and hence   can  be   observed to
distances of hundreds of Mpc  \cite{tammsand,hamuy96, rpk}.  They have
a  narrow  range in maximum  luminosity  and empirically an additional
relation exists between the luminosity of the supernova at its maximum
and  the  rate  at which   the   supernovae subsequently decreases  in
brightness,  \cite{phil,rpk}.  Bright  supernovae decline more slowly.
Using this correlation, the dispersion for type Ia supernovae drops to
about 12\% in luminosity, corresponding to an uncertainty of about 6\%
in the distance for a single supernova \cite{rpk}.  Currently no other
secondary  distance indicator rivals  this precision.   Unfortunately,
the exact mechanism for the ignition of the explosion has not yet been
theoretically or observationally established,  nor are the progenitors
known   with   any  certainty.    Ultimately,  confidence    in   this
empirically-based method will be strengthened as the theoretical basis
is more firmly established.

\subsection{The Tully--Fisher relation}

For spiral  galaxies,     the total (face--on)   luminosity   shows an
excellent correlation with the maximum rotation velocity of the galaxy
\cite{tf,ahm,pt,giovanelli97}.  This relationship  reflects the   fact
that more massive (and luminous) galaxies  must rotate more rapidly to
rotationally  support themselves.    Independent of distance,   galaxy
rotation rates can  be measured spectroscopically (from Doppler shifts
of  spectral features  of  hydrogen at  radio or optical wavelengths).
This  relation has  been  measured for   hundreds of   galaxies within
clusters,   and in the   general   field.  Empirically,   it has  been
established that the dispersion in this relation amounts to about 30\%
in    luminosity, or a  15\%    distance uncertainty for an individual
galaxy.  By measuring  a couple of dozen  or more galaxies in a single
cluster, the statistical uncertainty in  distance can be reduced to  a
few percent.

\subsection{Fundamental Plane}

For elliptical  galaxies, a correlation   between the stellar velocity
dispersion  and  the intrinsic luminosity    exists, analogous to  the
relation   between rotation   velocity  and   luminosity for   spirals
\cite{fj}.  Elliptical galaxies  also occupy a `fundamental plane'
wherein  the  galaxy  size  is   tightly correlated with   the surface
brightness       and     velocity    dispersion  of       the   galaxy
\cite{7S,djor,jorgen96}.  The scatter    in  this  relation is    only
10--20\%  in  distance.  Both the   Tully-Fisher and fundamental plane
relations will  be limited in precision  as distance indicators to the
extent that  the mass--to--light ratios  of galaxies are not universal
and that  star formation  histories may  vary   (that is, the  stellar
populations  within  galaxies have   different  mean ages or  chemical
compositions for  a given   mass).   Empirically, however,  with   few
exceptions, deviations from these  relations are  measured to be  very
small, providing compelling  evidence that mass--to--light and stellar
population variations are quantitatively constrained by the scatter in
the observed relations \cite{giovanelli98,jorgen96}.

\subsection{Surface Brightness Fluctuations}

Another  method with high internal   precision, developed by Tonry and
Schneider \cite{ts},  makes use  of the fact   that the resolution  of
stars within galaxies is  distance dependent.  In  each pixel on a CCD
detector, a given number of stars contributes  to the luminosity.  The
Poisson fluctuations from pixel to  pixel then depend on the  distance
to the  galaxy.  They have been  empirically determined to be a strong
function of the color of the stars.   Once other sources of noise (bad
pixels on the detector,  objects   such as star clusters,   background
galaxies, foreground  stars) have been removed,  by normalizing to the
average flux,  this  method  provides  a means of  measuring  relative
distances to galaxies that has been established empirically to yield a
precision of $\sim$8\% \cite{tonry97}.   With HST, this method is  now
being applied  out to velocities of  about 5,000 km/sec \cite{lauer98,
tonry97}.  This method is applied to elliptical galaxies or to spirals
with prominent bulges.

\subsection{From Relative to Absolute Distances}

The secondary methods  described  above (type  Ia supernovae, the  Tully-Fisher relation,   the fundamental plane,   and
surface brightness fluctuations)  provide  several means  of measuring
{\it relative}  distances to galaxies.   The absolute  calibration for
all  of these  methods is  presently  established   using the  Cepheid
distance scale.  To  give a specific  example, absolute distances  for
supernovae  require both measurements  of the apparent luminosities of
distant supernovae (the quantity observed),  as  well as distances  to
nearby galaxies in  which type Ia  supernovae have also been observed.
The distances to nearby type Ia supernovae galaxy  hosts are needed to
provide the absolute  luminosities of supernovae.    Only then can  an
absolute distance   scale  be set  for  the   more distant supernovae.
Although references  are occasionally made  to the  ``Cepheid distance
scale'' and the  ``supernova distance scale'',  the supernova distance
scale is  not independent of, but is  built upon, the Cepheid distance
scale.  With the exception of  theoretical  models of supernovae,  all
H$_0$ measurements  of   supernovae are calibrated using   the Cepheid
distance scale.   The  same holds true  for  all of the other  methods
listed above.

\subsection{\bf Systematic Effects in Distance Measurements}

Many distance indicators have sufficiently small scatter that with the
current numbers of  Cepheid calibrators, the statistical  precision in
their   distance  scales  is  5\%   or better.   The total uncertainty
associated with  the   measurement of distances   is higher,  however,
because of complications due  to other astrophysical effects.  Many of
these  systematic effects  are common  to all  of these  measurements,
although their cumulative impact may vary from method to method.

Dust grains in  the regions between stars, both  within our own Galaxy
and in external galaxies, scatter blue light more than red light, with
a  roughly   1/$\lambda$   dependence.    The  consequences   of  this
interstellar dust are two-fold: 1) objects become redder (a phenomenon
referred  to as reddening) and  2)   objects become fainter  (commonly
called extinction).  If no correction is made for dust, objects appear
fainter  (and therefore apparently   farther) than they actually  are.
Since the effects of  dust  are wavelength dependent,  corrections for
reddening and  extinction can be made if  observations are made at two
or  more  wavelengths   \cite{wlf1613,mf91,rpk,tonry97}.

A  second  potential  systematic   effect  is  that  due   to chemical
composition or metallicity.  Stars   have  a range  of  metallicities,
depending on the amount of processing by previous generations of stars
that  the gas  (from which they   formed) has undergone.  In  general,
older stars have  lower metallicities, although there is  considerable
dispersion at  any given age.  Metals  in the atmospheres of stars act
as an   opacity  source to  the radiation   emerging  from the nuclear
burning.  These  metals absorb   primarily  in the  blue part   of the
spectrum, and the radiation  is thermally redistributed  and primarily
re-emitted at longer (redder) wavelengths.

For any given method, there may also be systematic effects that are as
yet unknown.   However,  by comparing several  independent  methods, a
limit  to the total systematic error  in H$_0$ can  be quantified.  In
the next section, I turn back to the  measurement of Cepheid variables
and the  absolute  calibration of  the  extragalactic  distance scale,
reviewing recent progress both from the ground and from HST.

\section{Cepheid Distances to Galaxies}

\subsection{Recent Progress}

Significant progress  in the application of  Cepheid variables  to the
extragalactic  distance  scale has been made  over  the past couple of
decades \cite{jacoby92,mf91,fmtrimble}.   The areas where the  most
dramatic improvements have  been    made include the  correction   for
significant  (typically  0.5  mag)   scale  errors  in   the   earlier
photographic   photometry,    observations of   Cepheids   at  several
wavelengths,  thus   enabling corrections  for interstellar  reddening
\cite{wlf1613}, and empirical tests for   the effects of   metallicity
\cite{fm90,sasselov,kochanek,kennicutt}.  While dramatic progress  has
been made, both  from the ground and  with HST, there  is still a need
for further work, particularly regarding the zero point of the Cepheid
period-luminosity relation, as  well as in establishing accurately the
dependence of the period-luminosity relation on metallicity.



The practical  limit  for measuring  a well-defined  period-luminosity
relation from the   ground is only a   few megaparsecs.   Most of  the
Cepheid  searches before  the launch of  HST were  confined to our own
Local  Group of  galaxies  and the  nearest surrounding groups  (M101,
Sculptor and M81   groups)  \cite{mf91, jacoby92}.   Pre-HST,   only 5
galaxies with  well-measured  Cepheid distances provided the  absolute
calibration of the Tully-Fisher   relation \cite{fre90}, and a  single
Cepheid  distance, that  to  M31,  provided  the  calibration for  the
surface-brightness fluctuation   method   \cite{tonry91}.  It is   worth
emphasizing that before    HST,  {\it no}  Cepheid   calibrators  were
available for type Ia supernovae.

\subsection{ The HST H$_0$ Key Project: A Brief Description}

Broadly  speaking,   the  main aims  of the    HST  H$_0$ Key  Project
\cite{kfm95,fmmk98} were  twofold: first,  to  use the high  resolving
power of HST  to establish  an  accurate local extragalactic  distance
scale   based on the primary  calibration   of Cepheid variables,  and
second, to determine H$_0$   by  applying the Cepheid calibration   to
several secondary  distance  indicators operating further  out  in the
Hubble flow. The   motivation,  observing  strategy, and results    on
distances  to galaxies  have been described  in   detail elsewhere and
references  can be found in  the  above-cited references. Here a brief
summary is given.


As part of the HST H$_0$ Key  Project, Cepheid distances were obtained
for 17 galaxies  useful for the calibration  of  secondary methods and
determination   of H$_0$.  These   galaxies  lie at distances  between
approximately 3 and 25 Mpc.  They are located in the general field, in
small groups (for example, the M81 and the Leo I groups at $\sim$3 and
10 Mpc, respectively),  and in major  clusters (Virgo and Fornax).  An
additional  target, the  nearby   spiral galaxy, M101,  was chosen  to
enable   a  test  of   the  effects  of  metallicity    on the Cepheid
period-luminosity relation.  In addition, a team  led by Allan Sandage
has used HST  to  measure Cepheid distances  to 6  galaxies,  targeted
specifically to  be useful for the  calibration  of type Ia supernovae
\cite{sandage96}.  Finally, an HST distance to a  galaxy in the Leo I group
was measured by Tanvir and collaborators \cite{tanvir95}.

In addition to the increase in the numbers of HST Cepheid calibrators,
tremendous  progress  has  taken place  in  parallel  in measuring the
relative  distances  to  galaxies using    secondary techniques.   For
example, Hamuy and   collaborators    have  discovered 29   type    Ia
supernovae, and measured their peak magnitudes  and decline rates over
the range  of 1,000 to over  30,000 km/sec \cite{hamuy96}.  Giovanelli
and collaborators  have  measured rotational line  widths  and I--band
magnitudes  useful for  the Tully-Fisher relation  for a  sample of 24
clusters   over the velocity  range of   about 1,000  to  9,000 km/sec
\cite{giovanelli97}.  The  fundamental  plane for elliptical  galaxies
has been  studied  in a sample of  11  clusters from 1,100   to 11,000
km/sec    \cite{jorgen96}.  And, in  an   application   of the surface
brightness     fluctuation    technique,     Lauer and   collaborators
\cite{lauer98} have used HST to observe a galaxy in each of 4 clusters
located between about 4,000 and 5,000 km/sec.

These secondary indicators  have been calibrated  as part of the H$_0$
Key  Project   (type       Ia  supernovae   \cite{gibson99},       the
surface-brightness  fluctuation   method     \cite{ferrarese99},   the
fundamental plane  or D$_n$-$\sigma$  relation for elliptical galaxies
\cite{kelson99}, and the  Tully-Fisher  relation \cite{sakai99}).   In
addition,    the   planetary   nebula  luminosity   function    method
\cite{jacoby97} extends over the  same range as  the Cepheids (out  to
about 20 Mpc), and it offers a valuable comparison and test of methods
that operate locally (Cepheids,  RR Lyrae stars, tip  of the red giant
branch  (TRGB))  and those  that operate  at intermediate  and greater
distances   ({\it   e.g.,}  surface-brightness fluctuations   and  the
Tully-Fisher relation).    The  database  of Cepheid   distances  also
provide a means for evaluating less well-tested methods; for instance,
the globular  cluster   luminosity  function  \cite{ferrarese99}.  The
constraints provided by these papers have been combined, and a summary
of the  H$_0$ Key Project results  and their uncertainties is given in
\cite{mould99,wlfapjkp}.

The results from these papers are combined in the  top panel of Figure
1, a  Hubble diagram of distance  (in  megaparsec) versus velocity (in
kilometers/second).  The  slope   of this diagram   yields the  Hubble
constant  (in units of  km/sec/Mpc).  In  this  figure,  the secondary
distances  have all   been   calibrated using   the new   HST  Cepheid
distances.  The Hubble  line plotted has  a slope of 71.
Two features are immediately apparent from Figure 1.  First, all four
secondary indicators plotted show excellent agreement. Now that
Cepheid calibrations are available for all of the methods shown here,
there is not a wide dispersion in H$_0$ evident in this plot.  Second,
although the overall agreement is very encouraging, and each method
exhibits a small, internal or random scatter, there are measuraable
systematic differences among the different indicators at a level of
several percent.


\begin{figure*}
\psfig{figure=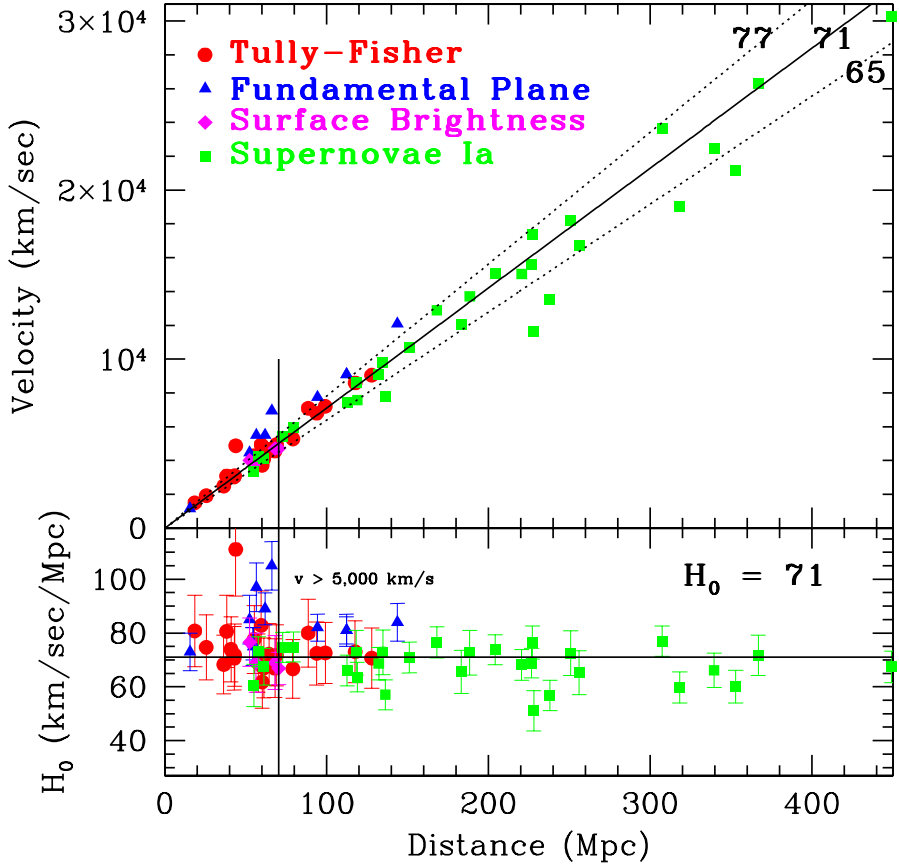,height=5in,angle=0}
\caption{
{\bf Top  panel}: A Hubble diagram of
distance versus velocity for  secondary distance indicators calibrated
by   Cepheids.   
Velocities in   this  plot  are corrected for a nearby flow
model \cite{mould99}.
microwave background (V$_{CMB}$)  reference frame.  
The symbols are as
follows:   Type Ia   supernovae --   squares,  Tully-Fisher  clusters
(I--band observations) --   solid circles, Fundamental  Plane clusters --
triangles,    surface brightness fluctuation
galaxies -- diamonds.  
A slope of H$_0$  = 71 is shown, flanked by $\pm$10\% lines.  Beyond
5,000  km/sec  (indicated by the    vertical   line), both   numerical
simulations and   observations suggest that  the   effects of peculiar
motions  are small.   The type Ia  supernovae  extend to  about 30,000
km/sec and the Tully-Fisher  and Fundamental Plane clusters  extend to
velocities of about  9,000 and 15,000  km/sec, respectively.  
However, the  current limit
for   surface  brightness fluctuations  is about  5,000  km/sec.  The latter
observations were obtained using new distances to galaxies in clusters
using HST \cite{lauer98}.
{\bf Bottom panel:} Residuals in H$_0$ as a function of velocity. 
}
\end{figure*}

The largest sources  of uncertainty in these individual determinations
of  H$_0$ include the numbers of  Cepheid calibrators  per method, the
effects  of metallicity, and the  velocity field on large scales. Each
method is impacted differently by each of  these factors. However, one
source of systematic  uncertainty,  that affects  {\it all}  of  these
methods, is  the  uncertainty in the adopted   distance to  the  Large
Magellanic Cloud.  This  nearby  galaxy provides the  fiducial Cepheid
period-luminosity relation    for the  Cepheid  distance  scale.   The
1-$\sigma$ uncertainty in the LMC  distance amounts to about  $\pm$7\%
\cite{wlfnobel,  westerlund,mould99}.  A second  source of  systematic
uncertainty  common to all methods  is the  photometric calibration of
HST magnitudes. Currently, this uncertainty is found  to be $\pm$ 0.09
mag (1-$\sigma$) \cite{mould99}.

The results for the   different secondary distance methods  have  been
combined in  several  ways to  determine  an overall value   for H$_0$
\cite{mould99,             wlfapjkp}.           These          results
\cite{gibson99,ferrarese99,kelson99,sakai99} are listed in Table    1.
For each method, the formal  statistical and systematic  uncertainties
are given.      The   systematic errors  (common   to    all of  these
Cepheid-based calibrations) are listed at  the end of  the table.  The
dominant uncertainties  are in  the  distance to    the  LMC and   the
potential effect  of metallicity on the Cepheid  PL relations, plus an
allowance is made for the  possibility that locally the measured value
of H$_0$ may  differ from the global value.   Also included is a  term
for systematic  errors in the calibration of  the HST photometry.  The
combined results  yield  H$_0$ =  71   $\pm$ 3  (statistical) $\pm$  7
(systematic) \cite{wlfapjkp}.

Because these determinations have a relatively small range (H$_0$ = 68
to 78 km/sec/Mpc), ultimately, there is good agreement in the combined
values of  H$_0$,  regardless of which   method is used.  In  one case
\cite{mould99}, the weights for  combining the various values of H$_0$
are determined using a  numerical, random-sampling strategy.  Each  of
the errors for these methods are treated as Gaussian distributions and
these   distributions are  randomly sampled    10$^5$  times.  A  more
realistic non-Gaussian probability distribution   for the distance  to
the LMC has  also been considered.   Based on this strategy, the value
of H$_0$ is found to be 71 $\pm$ 7 km/sec/Mpc, where no distinction is
made between  random   and systematic errors.  These  results   are in
excellent    agreement with   a   Frequentist   and Bayesian  analysis
\cite{wlfapjkp}.

\bigskip

\begin{table*}
\caption{\bf  SUMMARY OF KEY PROJECT RESULTS ON H$_0$}  

\begin{center}
  \begin{tabular}{lc} 
 \hline 
            &         \\
        \bf Method & H$_0$  \\[3pt] 
\hline 
 & \\
        \bf Local Cepheid galaxies & \bf 73 $\pm$ 7 $\pm$ 9  \\ 
 & \\
       \bf SBF & 69 $\pm$ 4 $\pm$ 6  \\
       \bf Tully-Fisher clusters & 71 $\pm$ 4 $\pm$ 7  \\ 
       \bf FP / D$_N -\sigma$ clusters & 78 $\pm$ 7 $\pm$ 8  \\ 
       \bf Type Ia supernovae   & 68 $\pm$ 2 $\pm$ 5  \\
       \bf  SNII   & 73 $\pm$ ~7 $\pm$ 7  \\ 
              &   \\
       \bf Combined   &   71 $\pm$ ~3 $\pm$ 7  \\
              &   \\
	\rm
     
       {\bf Systematic Errors} & {\bf $\pm$ 5 ~~~~~$\pm$ 3 ~~~~~$\pm$ 3  ~~~~~$\pm$ 4} \\
              & ~(LMC) ~([Fe/H]) ~(global) ~(photometry) \\

              &   \\
\hline

  \end{tabular}
\end{center}
\end{table*}

\medskip
 
\section{Remaining Issues}

\subsection{\bf Distance to the Large Magellanic Cloud}

It has become standard   for extragalactic Cepheid distances to  adopt
the Large  Magellanic   Cloud  (LMC) period-luminosity   relations  as
fiducial.  For the Key  Project as well as  the Sandage and Tanvir HST
studies, a distance modulus to the LMC of 18.5  (50 kpc)  mag has been
adopted for the zero point.

Although the factors-of-two  discrepancies in the distances  to nearby
galaxies have now  been eliminated, the  largest remaining uncertainty
in the distances  to galaxies remains  the absolute calibration.   For
example,  it  has  been  emphasized for  some  time   that   there are
disagreements  in  the {\it zero  points}  of the Cepheid  and some RR
Lyrae calibrations at a level of 0.15-0.3 mag (8 - 15$\%$ in distance)
\cite{fmvict93, wlfprinceton,  vandenB95}.  While  the Cepheid  and RR
Lyrae  distances agree to within  their stated errors, the differences
are systematic  (in the sense that the  RR Lyrae distances are smaller
than the Cepheid distances)  \cite{Walker92,Saha92}.  More recently, a
relatively new technique    for measuring nearby distances   to nearby
galaxies based  on a Hipparcos calibration of   the ``red clump'' have
also led to a smaller distance for the  LMC \cite{stanek98}.  However,
measurements of the  distance to M31 using the  red clump, the tip  of
the red giant  branch, and Cepheids  yield extremely good agreement at
24.47,  24.47 and 24.43 mag,  respectively.  It is  not yet understood
why  there is such good agreement  in M31 and not  in the LMC.  A very
recent  rotational parallax measurement   of masers in  the galaxy NGC
4258 also  supports a  shorter distance  scale \cite{herrnstein,maoz}.
However, recent measurements of the expanding ring for supernova 1987A
lead to  values of  the LMC distance   that range from 18.37 to  18.55
\cite{goulduza,panagia98} mag.



The  distance to the  LMC has been  reviewed recently  by  a number of
authors \cite{westerlund,walker99}.   The distribution of LMC distance
moduli is not Gaussian, and the range  is large, spanning 18.1 to 18.7
mag, with a  median of 18.45 and  68\% confidence  limits of $\pm$0.13
mag \cite{wlfnobel,mould99}.  The situation  is not  very satisfactory
as it stands,  since,  the largest  remaining component of  the  error
budget  for the Key  Project  is due to  this  uncertainty in the  LMC
distance. Note that  if the zero  point of the Cepheid  distance scale
was   adjusted by 0.2-0.3 mag    consistent with the shorter  distance
scale, the value of H$_0$ would be {\bf increased} by 10-15$\%$.


\section{\bf Does the Measured Value of H$_0$  Reflect the True, Global Value? }

Variations in  the  expansion rate  due  to peculiar velocities are  a
potential source  of systematic error  in measuring  the true value of
H$_0$.  For an accurate determination of H$_0$,  a large enough volume
must be observed to provide  a fair sample of  the Universe. How large
is large enough?

This  question  has  been addressed  quantitatively  in  a  number  of
studies.   Given   a model for  structure  formation,  and therefore a
predicted power  spectrum for density fluctuations, local measurements
of  H$_0$ can be compared with  the global value of H$_0$ \cite{tco92,
shiturn97,wang98}.  Many  variations  of cold dark  matter models have
been investigated,  and issues of both  the required volume and sample
size for the distance indicator  have been addressed.  The most recent
models   predict   that    variations     in   H$_0$    (that      is,
$<(\delta$H/H$_0)^2>^{1/2}$))   at the level   of   1--2\% are to   be
expected for the current  (small) samples of  type Ia supernovae which
probe out  to 40,000 km/sec, whereas  for methods  that extend only to
10,000 km/sec,  for  small samples, the variation  is  predicted to be
2--4\%.


Large density fluctuations will produce not  only variations in H$_0$,
but  a large   observed  dipole velocity with  respect  to  the cosmic
microwave  background radiation.   Cosmic Background  Explorer  (COBE)
measurements   of     our     dipole   velocity  of    627      km/sec
\cite{kogut93,fixsen94} have also been used to provide a constraint on
possible  variations in H$_0$,   completely independent of any assumed
shape for the underlying power spectrum for matter \cite{wang98}. This
constraint limits variations in H$_0$ on scales of 20,000 km/sec to be
less than 10\% (95\% confidence).

The overall conclusion from these studies is that uncertainties due to
inhomogeneities     in   the   galaxy   distribution   likely   affect
determinations of  H$_0$  at the few  percent level,  and this must be
reflected in  the total uncertainty  in  H$_0$.  However, the  current
distance indicators  are  now being   applied over sufficiently  large
depths  and angles that gross  variations  are statistically extremely
unlikely.   These  constraints   will tighten  as   larger numbers  of
supernovae are discovered,    and when all--sky measurements  of   the
cosmic  microwave background anisotropies are  made at smaller angular
scales.


\section{\bf The Age of the Universe }

\subsection{\bf Expansion Age}


Calculation of the expansion  age  of the  Universe requires  not only
knowledge of  the expansion rate, but  also knowledge of both the mean
matter    density    ($\Omega_m$) and    the  vacuum   energy  density
($\Omega_\Lambda$).   The force of gravity slows  the expansion of the
Universe. Hence, the higher the mass density, the faster the expansion
in the  past  would have  been  relative to  the present.   Until very
recently,  strong  arguments were  advanced to  support a cosmological
model   with  a    critical mass  density  $\Omega_m$  =    1,  and
$\Omega_\Lambda$ = 0  \cite{turner90,coles94}.
In this  simplest (the Einstein-de  Sitter)  model, the expansion age,
t$_0$ = 2/3 H$_0^{-1}$ is 9.3 Gyr $\pm$ 0.9 Gyr for a (round number) value of
H$_0$ = 70 $\pm$  7 km/sec/Mpc.  In  recent years, however, increasing
evidence suggests that the  total matter density   of the Universe  is
less than ($\sim$ 20--30\% of) the critical density \cite{neta}.
For H$_0$ = 70
$\pm$ 7   km/sec/Mpc, $\Omega_m$   = 0.3,  the  age of    the Universe
increases  from 9.3  to t$_0$  =  11.3  Gyr.   The effect  of  different
$\Omega$ values on the expansion age is shown in Table 2.  The errors
in the age reflect a 10\% uncertainty in H$_0$ alone.

In the past year, new data on type Ia  supernovae from two independent
groups  have provided evidence  for  a non-zero vacuum energy  density
corresponding  to       $\Omega_\Lambda$     =    0.7  
\cite{riess98},
\cite{perlm}.  If confirmed, the implication of  these results is that
the deceleration of the Universe due to gravity is progressively being
overcome  by a  cosmological constant term, and that the
Universe is in fact  accelerating   in its expansion.   Allowing   for
$\Omega_\Lambda$ =  0.7, under the assumption  of a flat ($\Omega_m$ +
$\Omega_\Lambda$ = 1)   universe,   increases the  expansion age   yet
further to t$_0$ = 13.5 Gyr.

\begin{table*}
\caption{Ages for Different Values of Cosmological Parameters}
\begin{center}
\begin{tabular}{l l l l}
\hline 
H$_0$    & $\Omega_m$     &  $\Omega_\Lambda$      &t$_0$ (Gyr) \\
\hline 
70       & 0.2      & 0        &  12 $\pm$ 1     \\
70       & 0.3      & 0        &  11 $\pm$ 1     \\
70       & 0.2      & 0.8      &  15 $\pm$ 1.5     \\
70       & 0.3      & 0.7      &  13.5 $\pm$ 1.5     \\
70       & 1.0      & 0        &  9  $\pm$ 1     \\
\hline 
\end{tabular}
\end{center}
\end{table*}

\subsection{\bf Other  Age  Estimates} 

Several methods exist for determining a minimum age  for our own Milky
Way  galaxy.  These ages provide an  independent check on cosmological
models,  since they provide   a hard lower   limit  to the age  of the
Universe.

A  firm  lower limit to  the  age of the  Galaxy  can be obtained from
radioactive dating of  isotopes produced  in stars  \cite{schramm}.
The Universe must  be  even older than   this
limit, of course, since we  know that the Galaxy did  not form all  of
its stars  in a  single burst.  Less  certain,  however, is  the exact
history of star  formation in the  Galaxy.  Models of galaxy evolution
include assumptions about the initial distribution of masses of stars,
the  rate at which   star  formation has  taken  place, and  how  much
processed  material   is ejected    from   stars  and  back  into  the
interstellar medium  for  reprocessing  through later   generations of
stars.  For  different   assumptions,  the  age estimates    for  this
particular  technique range   from    10 to 20    Gyr  \cite{schramm,
truran}.

The white dwarfs in our Galactic disk provide another means of putting
a lower limit on the  age of the  Universe.  These degenerate  objects
cool very  slowly; and so by  observing the coolest (and  faintest) of
these stars, models which  predict their cooling  rates can be used to
estimate a minimum age  of that population,  and therefore that of the
disk of the   Galaxy  \cite{oswalt96,bergeron}.  This lower  limit  is
found to be in the range of about 6.5 to 10 Gyr.

To   date,  the most accurate age    estimates are obtained  for stars
located in the globular   clusters in our   Galaxy.  For most   of the
lifetime of ordinary stars, hydrogen burns  into helium in the central
core,  and a balance   between the force of   gravity and  the outward
pressure  of radiation is established.    This phase  of evolution  is
referred to as the  ``main sequence''.  When  the hydrogen in the core
is exhausted, the  star leaves the main  sequence,  and the luminosity
and surface  temperature of the  star begin to increase  and decrease,
respectively.  By observing this  ``turnoff'' from the  main sequence,
and comparing to models of  stellar evolution, the  masses and ages of
stars in these systems can be estimated.


To  interface between  the   predicted,  model luminosities  and   the
observed, apparent luminosities of stars in globular clusters requires
accurate distances. Accurate distances are  needed not only for Hubble
constant  measurements, but   also  for  globular cluster ages.     In
addition,  corrections for reddening by  dust must again  be made, and
high-precision  chemical  abundances   measured.   The  importance  of
accurate  distances in this context  cannot be overemphasized.  A 10\%
error  in the distance to the  cluster results in a  20\% error in the
age of the cluster \cite{renzini}. (A 10\% error in the distance results
in a 10\% error in H$_0$.)

For the past approximately  30 years, the  calculated ages of globular
clusters     remained  fairly  stable     at  approximately   15   Gyr
\cite{vdb,chaboyer96,demarque}.  However,     new  results  from   the
Hipparcos  satellite  have led to  a  significant downward revision of
these  ages to  11-14 Gyr  \cite{reid,chaboyer98,pont}.  The Hipparcos
results,  in   addition to new  opacities  for  the  stellar evolution
models, have  provided parallaxes for relatively  nearby  old stars of
low  metal composition (the  so-called subdwarf stars), presumed to be
the nearby analogs of the  old, metal-poor stars in globular clusters.
Accurate distances to these stars  provide a fiducial calibration from
which the absolute  luminosities  of   equivalent stars in    globular
clusters  can  be determined  and  compared  with  those  from stellar
evolution models.

\subsection{\bf Is there an Age Discrepancy?} 

As  we have seen   above, in  a  low  matter-density universe with  no
cosmological  constant, H$_0$ = 70   results  in an  expansion age  of
$\sim$ 11-12 Gyr.    To within the  current  1-$\sigma$ uncertainties,
this  timescale  is comparable to   the most recent age  estimates for
globular clusters from Hipparcos.  The  absolute globular cluster ages
are uncertain at a level of about 2 Gyr.  It is also necessary to keep
in mind, however, that the age to compare  with the expansion age must
include also  the time required  for globular  cluster formation after
the Big Bang.   Generally, this timescale has been  assumed to be less
than 1  Gyr. 



To calculate  the total   uncertainty in  the expansion  age  requires
knowing  the uncertainties not  only in H$_0$,   but also in the other
cosmological  parameters.  At  the present time,   we do not know  the
matter density to 10\% precision.   The simplest statement that can be
made is that, to within the current  uncertainties, the expansion ages
are  consistent  with  the globular  cluster ages  either  for an open
universe or for a flat universe with non-zero $\Omega_\Lambda$.  For a
low density universe, with H$_0$ = 70 and the current uncertainties in
the globular ages,  one does not  require a cosmological constant, but
the remaining tension   between  the expansion and  globular   cluster
estimates is ameliorated if such a term is included.


Why is the discrepancy in ages  apparently no longer a serious problem
at the  present time?  Several  factors   have changed recently:  more
precise estimates  of  H$_0$ and  t$_0$   are now   available, and  in
addition,  current   observations    do  not  support    the  earlier,
theoretically-favored Einstein- de Sitter model  (with $\Omega_m$ = 1,
$\Omega_\Lambda$ = 0).  In fact, the better agreement in the expansion
and  globular cluster timescales discussed above   results not so much
from a change in H$_0$, (for H$_0$ = 70, an Einstein - de Sitter model
still yields an expansion age of 9  Gyr compared to  8 Gyr for H$_0$ =
80 \cite{freedman94}), as to the decrease in  the globular cluster ages
due to Hipparcos, and the increasing evidence for a low matter density
universe.

In Figure 2, the dimensionless product of $\rm H_0t_0$ is plotted as a
function   of $\rm \Omega$.  Two  different  cases are illustrated: an
open $\rm \Omega_\Lambda$ = 0 universe, and a  flat universe with $\rm
\Omega_\Lambda + \Omega_m$ = 1.  Suppose that both $\rm H_0$ and $t_0$
are both  known to $\pm$10\%   (1-$\sigma$, {\it including  systematic
errors}).  The dashed   and dot-dashed lines  indicate  1-$\sigma$ and
2-$\sigma$ limits, respectively  for values of  H$_0$ = 70  km/sec/Mpc
and t$_0$ =   12 Gyr.  Since the two   quantities H$_0$ and t$_0$  are
completely independent, the two  errors have been added in quadrature,
yielding  a  total  uncertainty on the   product  of $\rm  H_0t_0$  of
$\pm$14\% $rms$.     These values of  $\rm   H_0$  and $\rm  t_0$  are
consistent with   a universe  where   $\rm \Omega_\Lambda  \sim$  0.6,
$\Omega_m$ =   0.4.  Alternatively, an  open  universe  with $\Omega_m
\sim$ 0.2 is equally consistent.  For these values of H$_0$ and t$_0$,
the Einstein-de Sitter  model  ($\Omega_m$ =1, $\Omega_\Lambda$=0)  is
(marginally) inconsistent  at the  1.5$\sigma$ level. For  comparison,
there is an analogous plot to Figure 2 with H$_0$ = 70, but t$_0$ = 15
Gyr  \cite{wlftexas96}. 

Despite the enormous  progress  recently in  the measurements of  $\rm
H_0$ and $\rm  t_0$,  Figure 2  demonstrates that significant  further
improvements are still needed.  It is  clear from this figure that for
$\rm   H_0$ = 70 km/sec/Mpc,   accuracies of significantly {\it better
than} $\pm$ 10\% are required  to rule in or out  a non-zero value for
$\Lambda$.    

\begin{figure*}
\psfig{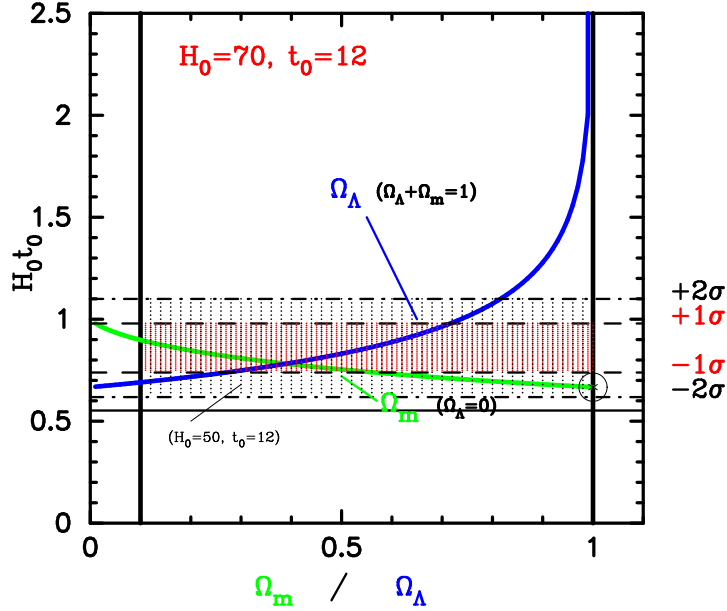}
\caption{H$_0$t$_0$ versus  $\Omega$ for H$_0$ = 70 km/sec/Mpc,
t$_0$ = 12 Gyr, and uncertainties of $\pm$10\% adopted for both ages.  
The dark line indicates the case of  a flat Universe
with $\rm \Omega_\Lambda + \Omega_m$  = 1.  The  abscissa in this case
corresponds  to  $\rm \Omega_\Lambda$.  The   lighter curve represents a
Universe with $\rm  \Omega_\Lambda$ = 0.     In  this  case, the abcissa
should  be  read as $\rm   \Omega_m$. The dashed  and dot-dashed lines
indicate 1-$\sigma$ and 2-$\sigma$ limits,  respectively for values of
H$_0$ =  70  km/sec/Mpc  and  t$_0$ =  12  Gyr in the case  where both
quantities are known to $\pm$10\% (1-$\sigma$).  The large open circle
denotes values  of  H$_0$t$_0$ = 2/3 and  $\Omega_m$  = 1 ({\it i.e.,}
those predicted by the  Einstein-de  Sitter model). Also shown
for comparison is a  solid line for  the  case H$_0$ = 50  km/sec/Mpc,
t$_0$ = 12 Gyr.}
\end{figure*}


\section{\bf Other Methods for Measuring H$_0$}

Ultimately,  for a    value   of H$_0$  and  its    uncertainty to  be
unambiguously established, it  is essential to have several techniques
that are based on completely different physics and assumptions.  There
are several methods for determining  H$_0$ that are independent of the
classical, extragalactic distance scale.  These  other methods offer a
number of advantages.  For  example,  the 3 methods  described  below,
based  respectively on the Sunyaev   Zel'dovich effect, time delays of
gravitational lenses,  and cosmic microwave  background  anisotropies,
all   can  be applied directly at    very large  distances, completely
independent of  the local extragalactic   distance scale.  However, to
date, the  numbers  of   objects,  or measurements for     these other
techniques is  still small, and  the internal systematics have not yet
been  tested  to the  same extent as   for the  extragalactic distance
scale.  Recently, there  has been progress in  all of these areas, and
ongoing and future experiments are likely to lead to rapid progress.

\subsection{Distances Based on the Sunyaev Zel'dovich Effect}

The underlying  principle  for  this technique  is similar   to  that
described for other distance indicators    in general: that is,    the
measurement of one  distance--dependent,  and one distance--independent
quantity. An excellent recent review of this subject has been
given by Birkinshaw \cite{birk99}.


As first described by Zel'dovich and Sunyaev  \cite{sz69}, some of the
low-energy cosmic microwave  background (CMB) photons from the surface
of last scattering  scatter off of the hot  electrons in the X-ray gas
in  clusters  and   generally gain    energy  through inverse  Compton
scattering.   As a  result,  measurements of  the microwave background
spectrum  toward rich clusters of galaxies   show a decrement at lower
frequencies (and a corresponding increase at higher frequencies).  The
size of the decrement thus depends on  the density of electrons in the
cluster and the path  length through the   cluster, but is  completely
independent of the cluster distance.  The observed X-ray flux from the
cluster is, however, dependent on the distance to  the cluster.  If it
can be assumed that the cluster is spherically symmetric, the distance
to the cluster can be solved for.



The greatest  advantages  of this method   are that it can  be applied
directly at large distances and   that it has an underlying   physical
basis.  However, there are a number  of astrophysical complications in
the   practical  application of  this method.    For  example, the gas
distribution in  clusters is not entirely  uniform: that is,  there is
clumping of the gas (which, if present,  would result in reducing $\rm
H_0$), there  are  projection effects (if   the clusters observed  are
prolate and   seen end on, the   true $\rm H_0$  could be  larger than
inferred).  Furthermore, this  method assumes hydrostatic equilibrium,
and a model for  the gas and  electron densities, and, in addition, it
is vital to eliminate potential contamination from other sources.  The
systematic errors incurred from all of these  effects are difficult to
quantify.

To date, a  range of values of $\rm  H_0$ have been published based on
this method ranging  from   $\sim$40 - 80   km/sec/Mpc  \cite{birk99}.
Two-dimensional interferometry maps of the  decrement are now becoming
available; the most    recent data for well-observed  clusters  yields
H$_0$ = 60 $\pm$ 10 km/sec/Mpc.
The  systematic  uncertainties are still large,  but  as more and more
clusters are observed,  higher-resolution X-ray maps and spectra,  and
Sunyaev--Zel'dovich maps,  become  available, the prospects  for  this
method are improving   enormously \cite{birk99,carlstrom}.  The
accuracy of this method will be considerably improved when a sample of
clusters has been identified independent of X-ray flux.

 
In Figure  3, a   Hubble  diagram of log d    (distance) versus log  z
(redshift) is shown. Included in this plot  are 4 clusters (Abell 478,
2142, 2256)  with cz $<$  30,000 (z$<$0.1) km/sec listed by Birkinshaw
\cite{birk99}   (his Table  7) as   being   clusters with reliable  SZ
measurements.  These data are overplotted with the Key Project Cepheid
and  the   secondary--method distances  shown in   Figure  1.  These 4
clusters  extend over  the same  current  range as type Ia supernovae.
Although  SZ measurements are available  out  to significantly greater
redshifts, beyond a redshift  of $\sim$0.1, the effects of  $\Omega_m$
begin to become significant.  No SZ clusters at z$>$0.1 are shown.  It
is encouraging to see how consistent the  results are over 2.5 decades
in redshift.  The local Cepheids (corrected for  the local flow field)
show more scatter, as expected. But  a value of  H$_0$ = 71 km/sec/Mpc
is consistent with all of the data shown, from  the local Cepheids out
to type Ia supernovae and the Sunyaev--Zel'dovich clusters.



\begin{figure*}
\psfig{figure=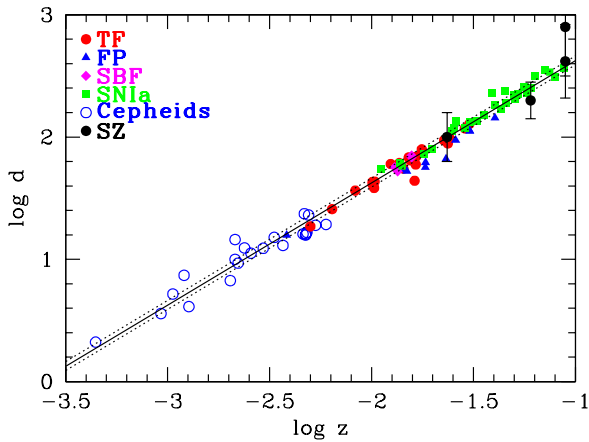,width=3.5in,height=3.5in,angle=0}
\caption{ Top panel: A local plus far-field Hubble diagram of distance
versus  velocity  including Cepheids,  secondary distance  indicators,
plus Sunyaev Zel'dovich  lens  measurements extending to  $\sim$30,000
km/sec. 
The nearby galaxy velocities have  been corrected for the local flow
field following Mould {\it et al.} \cite{mould99}.  The symbols are as
follows: Type  Ia supernovae --  squares, Sunyaev Zel'dovich method --
solid  circles  with their  published error bars   -- Tully--Fisher --
solid  circles,    fundamental  plane  --   solid  triangles,  surface
brightness fluctuations -- solid diamonds, Cepheids -- open circles.
A slope of H$_0$ =  71 is illustrated by  the solid line. Dashed lines
indicate  $\pm$10\%.  The scatter   in  the far-field measurements  is
still  significantly larger than for     type Ia supernovae, but   the
results   are  consistent to  within  the  current uncertainties.  The
prospects for decreasing the scatter in  the Sunyaev Zel'dovich method
appear very good in the near term  as higher resolution interferometry
and X-ray maps, for larger numbers of clusters become available.  }
\end{figure*}


\subsection{Gravitational Lenses}

A second   method  for  measuring H$_0$   at    very large  distances,
independent  of the need for   any local calibration,  comes from  the
measurement of   gravitational  lenses.  Refsdal  \cite{ref64,  ref66}
showed that a measurement of the time delay and the angular separation
for different images of a variable object such as a quasar can be used
to provide a measurement of $\rm  H_0$.  This method offers tremendous
potential not only  because it can be applied  at great distances, but
it is based on very solid physical principles \cite{blandnar}.


Difficulties with  this method  stem  from the  fact that astronomical
lenses are extended galaxies  whose underlying (luminous or dark) mass
distributions are not independently  known.  Furthermore, they may  be
sitting in more complicated group or cluster potentials.  A degeneracy
exists  between the mass  distribution of  the  lens and  the value of
H$_0$    \cite{schech97,romankoch}.    Ideally   velocity   dispersion
measurements as  a  function of position are   needed to constrain the
mass distribution of the lens.   Such measurements are very difficult,
but are recently  becoming available  \cite{tonryfranx}.  H$_0$ values
based  on  this  technique   appear  to  be converging   to   about 60
km/sec/Mpc,   although  a range   of   40  to  80  has  been published
\cite{schech97, romankoch,tonryfranx,impey98}.




\subsection{Cosmic Microwave Background Anisotropies }

The underlying   physics governing the shape   of the cosmic microwave
background  (CMB)   anisotropy  spectrum  can   be  described  by  the
interaction of a very tightly coupled fluid  composed of electrons and
photons  before recombination \cite{huwhite,sz70}.  If the underlying
source of  the   fluctuations   is  known,  the power     spectrum  of
fluctuations can be computed and compared with observations.  Over the
next few years,  increasingly more accurate  measurements will be made
of the fluctuations  in the CMB,  offering the potential to measure  a
number of cosmological parameters. This field is becoming increasingly
data rich  with a number  of planned and ongoing long-duration balloon
experiments, and planned   satellite experiments ({\it e.g.},  MAP and
Planck).  Using the  CMB data  in combination   with other data,   for
example,  the Sloan survey,  appears  to be a   promising way to break
existing model   degeneracies \cite{eisen}, and  measure   a value for
H$_0$.

\section{\bf The Future  }

A critical issue affecting the  local determinations of H$_0$  remains
the zero-point calibration of  the extragalactic distance scale  (more
specifically, the Cepheid  zero  point).  The  most promising  way  to
resolve this   outstanding uncertainty is  through accurate  geometric
parallax measurements.  New  satellite  interferometers are  currently
being planned  by NASA (the Space  Interferometry Mission  -- SIM) and
the European Space Agency (a mission known as GAIA) for the end of the
next decade.  These interferometers will be capable of delivering 2--3
orders of  magnitude  more  accurate parallaxes  than  Hipparcos ({\it
i.e.,} a  few  microarcsec astrometry),  reaching   $\sim$1000$\times$
fainter limits.  Accurate parallaxes for large numbers of Cepheids and
RR  Lyrae  variables  will  be  obtained.   Moreover, in   addition to
improving the  calibration  for the distance to  the  LMC, it will  be
possible  to measure rotational   parallaxes for several nearby spiral
galaxies, with distances accurate to a few percent.


Improvement to the    photometric  calibration for the   HST   Cepheid
measurements will   be possible with the  Advanced  Camera for Surveys
(ACS), currently  scheduled  to fly in  the  year 2000.    Next to the
uncertainty in  the distance  to   the  Large Magellanic Cloud,    the
photometric zero point   contributes  the  second largest  source   of
systematic error in the determination  of H$_0$. New ACS  observations
should quickly yield a higher accuracy than is currently possible with
the Wide Field and Plantary Camera 2 now in use.

\section{ Concluding Remarks}

Recent results on the determination of H$_0$ are encouraging.  A large
number  of  independent secondary methods  (including  the most recent
type  Ia supernova      calibration by    Sandage  and   collaborators
\cite{saha99})  appear to  be converging on  a value  of  H$_0$ in the
range of  60 to  80 km/sec/Mpc.   While   only a few  years  ago, some
published Cepheid distances to  galaxies \cite{fmvict93} and values of
H$_0$ differed by a factor of two , the $rms$ differences are now at a
level of 10\%. Given the historical difficulties in this subject, this
is welcome progress.  However, the need to improve the accuracy in the
determination of H$_0$ is certainly not over. For an $rms$ uncertainty
of 10\%, the  95\% confidence range restricts  the value of H$_0$ only
to  57 $<$  H$_0$ $<$ 85   km/sec/Mpc, underscoring the importance  of
reducing  remaining  errors in the  distance  scale ({\it  e.g.}, zero
point, metallicity).

Even though there has been considerable progress recently, the current
accuracy in H$_0$ is insufficient to discriminate between cosmological
models that are open and those that  are flat with non-zero $\Lambda$.
Before compelling constraints  can be made  on cosmological models, it
is imperative to rule out remaining sources of systematic error.  With
a  value of  H$_0$ accurate  to   10\% (1-$\sigma$) now  available, it
brings  into sharper focus smaller  (10-15\%) effects which used to be
buried in the noise in the era of factor-of-two discrepancies.


\bigskip
\noindent
ACKNOWLEDGMENTS ~~It is  a  pleasure  to  contribute to  this   volume
commemorating David  Schramm.  David was  extremely encouraging to me,
always rightfully skeptical,  but very interested  in the most  recent
observational results.  I  sincerely thank all  of my collaborators on
the extragalactic distance scale  over the past 15 years, particularly
B.  F.  Madore. In addition my thanks to all of the members of the HST
H$_0$ Key  Project team, whose enormous  contributions enabled the Key
Project to be undertaken:   R.  Kennicutt, J.R.  Mould  (co-PI's),  F.
Bresolin, L.  Ferrarese, H.  Ford, B.  Gibson, J.  Graham, M.  Han, P.
Harding,  J.  Hoessel,  J.  Huchra, S.    Hughes, G.  Illingworth,  D.
Kelson, L.  Macri, B.F.   Madore, R.  Phelps, A.   Saha, S.  Sakai, K.
Sebo, N.   Silbermann, P.   Stetson,  and  A.  Turner.   Some  of  the
results presented in this  paper  are based  on observations with  the
NASA/ESA Hubble   Space  Telescope, obtained  by  the  Space Telescope
Science  Institute,  which is   operated   by AURA,  Inc.   under NASA
contract  No.  5-26555.   Support for  this work was  provided by NASA
through grant  GO-2227-87A from STScI.  This  work has benefited from
the use of the NASA/IPAC Extragalactic Database (NED).





\end{document}